# The improbable event of spontaneous cell rejuvenation

Ian von Hegner
University of Copenhagen
Universitetsparken 13, DK-2100 Copenhagen, Denmark

**ABSTRACT** Unlike most other laws of nature, the second law of thermodynamics is of a statistical nature, according to Boltzmann, meaning that its reliability arises from the vast number of particles present in macroscopic systems. This means that such systems will lead towards their most likely state, that is, the one with the most homogeneous probability distribution. However, Boltzmann states that entropy-decreasing processes can occur (without doing any work)—it is just very improbable. It is therefore not impossible, in principle, for all 6 x $10^{23}$ atoms in a mole of a gas to spontaneously move to one half of a container; it is only fantastically unlikely. A similar idea has here been applied to a human cell. All somatic cells seem to age and deteriorate in unfavourable conditions. If the ageing process is defined as the accumulation of dysfunctional polymers resulting from, among other things, chemical bond breakage, where polymers aggregate into harmful arrangements, spreading out randomly in the cell and leading to an altered function, then it also applies that there will be a difference in entropy between, for instance, a 20-year-old individual and the same individual at age 80. The goal of this article is to demonstrate that the second law does not tell us that the cell necessarily must go towards a high entropy state and stay that way but that it is possible—according to statistical mechanics—for an old cell to experience a return to a younger state. We find the probability of this spontaneous return to a more ordered state to be expressed by P = $10^{-202 \times 10^{-889}}$. In spite of this number, it does show that a reversal of the ageing process is not prohibited by nature. There is a theoretical possibility of rejuvenation. Whether this will ever become a practical reality is another matter.

## INTRODUCTION

“Senescence” refers to a multidimensional state or process of physical and psychological ageing in an individual over time. Organism senescence is the accumulation of change in the physiology of an organism as it ages, especially after its maturity. Cellular senescence is a phenomenon in which isolated normal cells show a decreasing ability to divide and maintain complete functionality in a culture [9]. There are a number of explanations as to why senescence occurs. Only a few general theories attempt to explain why almost all living organisms age. Other theories are more modest in scope and address the question, How do we age? There are mainly two main groups of ageing theories: programmed ageing and stochastic ageing [2].

Programmed theories argue that ageing is programmed into the organism and occurs due to gene expression changes or intrinsic timing mechanisms, e.g., genetic timers. Stochastic theories argue that ageing is a result of damage caused by biological processes that accumulates over time. It occurs as the result of change or unfortunate events, e.g., free radical damage.

There is no question that many aspects of ageing look like the accumulation of damage. Examples are oxidative damage, mutations, and protein cross-linkages. It seems that ageing, in large measure, is a consequence of stress acting on the basic unit of life—the cell. Damage at the molecular level causes, as in machinery and other inanimate objects, the mechanisms of the cell to eventually malfunction. However, unlike inanimate objects, living beings possess the ability to replace, synthesize and repair dysfunctional molecules to maintain the biological activity of their molecules and retain individual life [9]. For most life forms, however, this balance slowly changes to a condition in which the accumulation of dysfunctional molecules starts to move beyond the capacity to repair them. After reproductive maturation, the mechanisms responsible for the repair capacity themselves start falling prey to the same type of damage as their substrate molecules, thereby further increasing the ageing process [9]. The essential mystery is why the body is able to avoid these problems for many decades but then allows the damage to occur in old age. These are reasons why ageing needs a more complex theory.

Some authors have suggested that ageing is a result of entropy, since entropy in a certain sense requires increasing disorder and therefore deterioration [4]. Entropy, as discovered by Ludwig Boltzmann [3], is a measure of dispersion for the concentration of free energy when unhindered—in some way, effectively a measure of disorder in a system. Therefore, the second law of thermodynamics can be stated as "The entropy of the universe tends to increase with time and can never decrease" [10]. Strictly speaking, the second law is certainly responsible for ageing as well as for any deterioration because the second law is an indication of the arrow of time. That is, without the second law, life would be in thermal equilibrium [11].

However, entropy increases inevitably within a closed system, and living beings are not closed systems. It is a defining feature of life that organisms are very far from equilibrium, and their lives are in constant flux between the increase of entropy they would experience alone and the upkeep they manage by taking advantage of free energy. They routinely counteract entropy by feeding on free energy in food taken from their surroundings to grow and unload their entropy as waste [8]. Therefore, while entropy is clearly responsible for overall ageing, this relationship is too general to be of any consequence in applied science.

A more specific theory is needed, one that also explains why cells overcome the effects of the second law only for a limited time and ageing occurs after all. Such a theory has been formulated by Toussaint et al. [13] as follows: "A theory of ageing considered as a multi-step process leading the cell through a sequence of defined stages characterized by a lower level of entropy production and finally to a critical level of errors involving cell death. One of the predictions of this model is that external stresses which can be considered as fluctuations would accelerate the evolution of the cell from one state to the other according to the intensity of the stress. Seven morphotypes have been observed in the serially cultivated human fibroblasts, cells passing progressively from one morphotype to the other. … All stages are not equally stable with morphotypes III and IV being the most stable. The positive effect on the increased shift of these cells from one morphotype to the other by two different stresses confirms one of the prediction of the thermodynamic model which states that cellular ageing can be considered as a multi-step process which can be speeded up by various external modifications."

This model transcends both the stochastic and programmed theories. Whether senescence as a biological process can be slowed down, halted or even reversed is an area of current scientific research. It is a useful practice in all types of engineering to first explore whether nature has already, in practice or in theory, accomplished a similar feat. Therefore, if ageing is a result of ever-increasing damage to the cell, would a reversal of the disorder be possible? That is, would a system of molecules be able to recover its previous order? This brings us to statistical mechanics.

## THEORY

The purpose of statistical mechanics is to explain the behaviour of systems comprised of a very large number of entities. It applies probability theory to the study of the thermodynamic behaviour of such systems. It delivers a framework for relating the microscopic properties of single atoms and molecules to the macroscopic properties of materials that can be observed in common life [10]. Statistical mechanics delivers a molecular-level explanation of macroscopic thermodynamic quantities such as work, heat, free energy, and entropy. The explanation offered by statistical mechanics of the asymmetry over time of processes has an important role in efforts to understand the asymmetries of causation and of time itself.

The idea that the second law of thermodynamics is about disorder is due to Boltzmann's view of the second law. Specifically, it was his effort to reduce it to a stochastic collision function, or law of probability, resulting from the arbitrary collisions of mechanical particles. Although this idea of disorder is not entirely correct, it is useful for the purposes of this article.[1] Boltzmann put forward a combinatorial definition of entropy [3]. According to this definition, a distribution of particles over finite cells is the number of particles in each cell, and a complexion is the specification for each particle of the cell to which it belongs. The likelihood of a given distribution was taken to be proportional to the corresponding number of complexions. Since all particles have the same probability in the distribution—the number, not the identity, of particles in a cell is relevant to the definition of a microstate—entropy is defined as the number of distributions compatible with a given macrostate [3]. In equilibrium, each microstate that the system might be in is equally likely to happen. This brings us to the conclusion that the second law will hold, on average, with a variation on the order of $1/\sqrt{N}$, where N is the number of molecules in the system. This means that the global macrostate emerges as not only the most unique state but also the state that is most likely to happen and the final state that any evolution will lead to, beginning from an arbitrary initial state.

Boltzmann modeled gas molecules as colliding billiard balls in a box and remarked that with each collision non-equilibrium velocity distributions would become increasingly disordered, leading to an end state of macroscopic uniformity and maximum microscopic disorder, that is, the state of maximum entropy. Therefore, the second law was a result of the fact that in a universe of mechanically colliding molecules, disordered states are the most probable. Thereby, Boltzmann made the second law a direct expression of the laws of probability: the entropy, S, is proportional to the logarithm of the probability of that state, $S = k_B \log(\Omega)$. Since there are so many more possible disordered states than ordered ones, a system will usually be either in the state of maximum disorder—the macrostate with the highest number of accessible microstates such as a gas in a box at equilibrium—or moving towards the state of maximum disorder. A dynamically ordered state, that is, one with particles moving 'at the same speed and in the same direction', said Boltzmann, is thus 'the most improbable case conceivable … an infinitely improbable configuration of energy' [3].

This definition of entropy implies many new and extremely interesting possibilities. The second law is a probability statement; it tells us the most probable event, rather than the only possible event. If sufficient time is provided, even the most improbable states can exist [11]. The reliability of the second law arises from the large number of particles present in macroscopic systems. For everyday situations, the probability that the law will experience improbable events is practically zero.

However, probabilities are not certainties; for systems with a modest number of particles, thermodynamic parameters, including entropy, may show significant statistical deviations from the values predicted by the second law [7]. Therefore, it is statistically possible for a system to achieve moments of non-equilibrium. For example, the law says that the molecules of a gas tend to disperse towards a uniform distribution and, yet, there is some non-zero (albeit, incredibly small) probability that the reverse process might take place. It is not impossible, in principle, for all $6 \times 10^{23}$ atoms in a mole of a gas to spontaneously move to one half of a large container and the overall entropy to fall; this event is only highly unlikely, not impossible. Certainly, such an event would demand very special circumstances; however, there seems to be a chance in which the second law could allow this, provided there is a sufficient amount of time or 'luck'.

[1] Occasionally, entropy is defined as disorder; however, this definition can be misleading. Strictly speaking, entropy is a measure of dispersion of energy or a measure of available microstates, which means that a system's entropy is a property that depends on the number of ways in which energy can be distributed among the molecules in the system. Disorder is a metaphor, rather than a definition, for entropy. Metaphors are beneficial only when they are not identical in all respects to what they describe. An increase in entropy results from a decrease in free energy, whereas an increase in disorder arises from isergonic phase dispersion (the loss of coherence in motions).

There are numerous ways in which the second law of thermodynamics can be stated, and one more correct definition states that the entropy of a thermally isolated system will almost never decrease spontaneously. Therefore, in principle, it is possible that a cup of coffee kept on a table becomes warmer as it draws in heat from its cooler surroundings. Although this is usually considered highly improbable, spontaneous reductions in entropy can and do occur on very small scales. Even such statistically improbable events where hot particles 'borrow' the energy of cold ones, such that the cold particles become colder and the hot hotter, for a moment, can be expected to occur from time to time in a vessel containing only a few particles. This has, in fact, been observed at a small enough scale whereby the probability of such an event happening is significant [15]. It is, however, still very unlikely that we would ever notice a spontaneous entropy reduction at the macro scale.

There has been much debate concerning Boltzmann's interpretation of the second law, but this discussion is not the objective in this context, and it will not be dealt with here. For the remainder of this article, we intend to accept his general premise and apply its consequences to the biological ageing process. A statistical mechanical approach can be used to study the biological ageing process and the highly improbable reversal of it. An ordered arrangement n, being less likely, represents a lower entropy configuration, and events that decrease the entropy of the system necessitate a change from more-random states to less-random states. This insight is useful when one analyses what happens for polymers, such as protein and DNA. However, first, we should build a foundation by exploring gas in a container.

**RESULTS**

If we put a drop of milk in a cup of coffee, it will spread out, but nothing in the laws of nature prevent the milk from coming together again in a drop. Boltzmann himself chose Thomson's example to demonstrate that we should not expect a mixture of nitrogen and oxygen gases to separate in a box after a month, with oxygen in one half and nitrogen in the other half of the container, although, from the viewpoint of probability theory, that event is highly improbable, but not impossible [3]. Let us illustrate this possibility by first examining the classic example of gas particles in a container. We will then proceed to apply similar thinking to a somatic cell.

GAS IN A TWO-COMPARTMENT CONTAINER At a microscopic level, we can easily understand why some processes happen of their own accord, while others do not. A spontaneous event corresponds to the rearrangement of particles from a less-probable situation to a more-probable one. A nonspontaneous event, by contrast, corresponds to movement from a probable situation to an improbable one [10]. An example of what probability has to do with a spontaneous event is given by the expansion of a gas into a vacuum. Let us estimate the likelihood that the process of gas expansion from compartment A into a connected compartment B will reverse itself, i.e., the probability that the gas particles will all gather again in compartment A. Assume that we begin from an isolated system in thermal equilibrium, then each of the $\Omega_i$, say, accessible states are equally probable. If we now remove some of the constraints that are imposed on the system, then obviously, all of the microstates that were formerly accessible to the system are still accessible, but a number of additional states will, in general, become accessible. Thus, removing constraints will have the effect of increasing, or possibly leaving unchanged, the number of microstates that are accessible to the system. If the final number of accessible states is $\Omega_f$, then we can write the following

$$\Omega_f \geq \Omega_i. \qquad (1)$$

Right after the constraints are removed, the systems in the ensemble will not be in any of the microstates from which they were previously excluded. Therefore, the systems only occupy a fraction as follows

$$P_i = \frac{\Omega_i}{\Omega_f} \quad (2)$$

of the $\Omega_f$ states that are now accessible to them. This is obviously not an equilibrium situation. In fact, if $\Omega_f >> \Omega_i$ then the configuration in which the systems are only distributed over the original $\Omega_i$ states is a very unlikely one. The ensemble will evolve in time until a more likely final state is obtained, in which the systems are evenly distributed over the $\Omega_f$ available states. As an example, consider a system that consists of a container that is divided into two compartments of equal volume. Suppose that, initially, one compartment is filled with gas and the other is empty. The constraint imposed on the system is, thus, that the coordinates of all of the molecules must lie within the filled compartment; that is, the volume accessible to the system is $V = V_i$, where $V_i$ is half the volume of the container. The constraints that are imposed on the system can be relaxed by removing the partition and allowing gas to flow into both compartments. The volume that is accessible to the molecules is now $V = V_f = 2V_i$. Right after the partition is removed, the system is in a highly improbable state. At constant energy, the variation of the number of accessible states of an ideal gas with the volume is as follows

$$\Omega \propto V^N \quad (3)$$

where N is the number of particles. If we select a particular molecule and designate it as number 1, we find that it is occasionally in compartment A and is occasionally in compartment B. Since the molecule's motion is arbitrary and the two compartments contain the same volume, the molecule should spend half its time in each. Thus, the probability of observing the state right after the partition is removed in an ensemble of equilibrium systems with volume $V = V_f$ is as follows

$$P_i = \frac{\Omega_i}{\Omega_f} = \left(\frac{V_i}{V_f}\right)^N = \left(\frac{1}{2}\right)^N. \quad (4)$$

The probability of locating molecule 1 in compartment A is, therefore, 1/2.[10] Next, let us study the probability that two molecules, which are designated as 1 and 2, are both in compartment A. There are four possible ways such that these two molecules can be arranged in the two compartments. All four are equally probable, but only one has both molecules in compartment A. Thus, there is one chance in four that molecules 1 and 2 are both in compartment A. The probability will be $p = (1/2)^2$ or 1 in 4, because each molecule will have a probability of 1/2 of being in compartment A, and they move independently. In general, the probability of all N gas molecules being in compartment A at once is $p = (1/2)^N$.

Let us study a simple system consisting of 10 molecules. We begin with a relatively ordered system, i.e., all 10 molecules are found in one-half of the container. As they move about and collide with one another, the molecules eventually distribute themselves throughout the entire container. However, one can easily imagine that by watching the random movement of the molecules, it might happen — by chance alone — that all of the molecules become segregated on one side. In fact, they all eventually find themselves in the same half of the container once again. While the system spends most of its time in the state in which the molecules are scattered throughout the entire container, once every few minutes, the molecules return briefly to a more ordered state. Given just 10 molecules, the system periodically experiences a spontaneous decrease in entropy. That this can be expected to happen from time to time can be deduced from the fluctuation theorem; thus, it is not impossible for the molecules to order themselves [7].

Let us study a similar situation for 100 molecules. For 100 independently moving molecules, the probability that the 50 fastest ones will be located in the left part of the container at any moment is $p = (1/2)^{50}$. Likewise, the probability that the remaining 50 slower molecules will be located in the right part at any moment is $p = (1/2)^{50}$. Therefore, the probability of finding this fast-slow separation as a result of arbitrary motion is the product $(1/2)^{50}$ $(1/2)^{50} = (1/2)^{100}$, which corresponds to approximately 1 in $10^{30}$. This works out to be approximately $10^{-30}$, which is a negligible quantity! Therefore, this time, spontaneous order would be highly unlikely. If the container contains in the order of 1 mol of molecules, then $N \sim 10^{23}$, and this probability is incredibly small: $P_i \sim \exp(-10^{23})$. If we had 1 mol of gas in the containers, where we take 12 grams of $CO^2$ molecules (approximately $6.022 \times 10^{23}$ molecules), then the probability p that at some later time, all of them have arranged themselves in compartment A at the same time would be as follows

$$P = \left(\frac{1}{2}\right)^{6.022 \times 10^{23}} \tag{5}$$

$$= \frac{1}{2^{6.022 \times 10^{23}}} = 1.54 \times 10^{-25},$$

which is an extremely improbable event! Because there are so many particles in a mole of gas (or any other macroscopic quantity), the probability that spontaneous expansion will reverse itself is tremendously small. The reversal is so improbable as to be impossible in any real situation. Clearly, the system will evolve towards a more probable state. The chances of all of the gas molecules in a container spontaneously bunching up in one end are not zero, but they decrease as the size of the system increases. This improbability gives rise to a statistical arrow of time [10].

Let us again study a system of 100 molecules that are free to bounce around arbitrarily in a container. If we increase the number of molecules in the container, the time during which the system stays in its disordered state increases to what seems to us to be almost an eternity. Estimates show that only once in every $1.5 \times 10^{22}$ years will all 100 molecules reverse back to one side of the container. If we estimated the probabilities for the real number of molecules in the container ($10^{23}$), we would basically never see the system spontaneously decrease its entropy. Therefore, in summary, to see significant spontaneous reductions in entropy, we need either a) very small systems or b) extremely large timescales [10].

## THE CELL AS A MULTI-COMPARTMENT CONTAINER

The cell is the basic structural and functional unit of all known living organisms. It is the smallest unit of life that is defined as a living entity. The cell carries out a huge number of biochemical reactions each minute and constantly transports essential molecules from place to place, takes in nutrients, expels waste, and reproduces new cells [1]. The components of cells are proteins, nucleic acids, carbohydrates, and lipids, and these are the four major molecules that constitute cell structure and take part in cell functions. For example, a tightly organized arrangement of lipids, proteins, and protein-sugar compounds form the plasma membrane. The organelles are built largely from proteins, and the DNA and RNA build the huge number of proteins that the cell needs [1]. Part of the cell might be seen, from a biophysical point of view, as a kind of quasi-crystal in which a given set of molecules occupies a diminished number of energy states. The cell membrane is, for instance, a form of crystal, namely, a lyotropic liquid crystal, and many proteins and even the cell genome can be considered to be a liquid crystal [6].

Imagine that instead of looking at a situation with a large number of air molecules in a two-compartment container, as in the previous section, we are now looking at a situation with a huge number of polymers in a cell. A probabilistic approach must still be used. How do we expect the polymers to spread out, all in one place or evenly throughout the cell? There is no law of physics demanding that it must be an even spread. Rather, it is a question of equal probability of all arrangements, i.e., a vast number of arrangements that correspond to even spreading.

In the following calculation, the cell is thought to be divided into a large number of small compartments or, more precisely, a thousand compartments (a gross oversimplification, but it will ease the calculations). Even spreading corresponds to the same number of polymers in each compartment; “all in one corner” corresponds to all the polymers being in a single compartment. We have the following data:[2] number of polymers: $1 \text{ x } 10^9$, cell size: $1 \text{ x } 10^{-9}$ cm³, size of a compartment: $1 \text{ x } 10^{-6}$ $\text{cm}^{3,}$ hence, the number of compartments is $1 \text{ x } 10^3$. The number of ways of placing all the polymers into a single compartment equals 1. Arranging the polymers evenly in each compartment is given by

$$\frac{1 \; x \; 10^9}{1 \; x \; 10^3} = 10^6, \tag{6}$$

and the number of ways of arranging the polymers evenly among all compartments is given by

$$(10^3)^{10^9}, \tag{7}$$

a number that without exaggeration can well be described as vast.

All somatic cells in eukaryotes seem to age and deteriorate in unfavourable conditions. One might think of aging as the accumulation of dysfunctional polymers resulting from, among other things, chemical bond breakage, where polymers aggregate into harmful arrangements, spreading randomly out in the cell, leading to a different product or biological inactivity. At a deeper level, it appears that cellular aging can be observed to be a multi-step process, where cells shift progressively through seven morphotypes, as characterized by a lower level of entropy production, meaning that the capacity of the cell to transform energy into work is falling over time, which eventually, leads to a critical level of errors involving cell death [14]. Vast changes in entropy are required to reorganize the large protein and DNA molecules that constitute functional cells. The probabilities for spontaneous reassembly of such polymers are very low, and no set of molecules can condense into a smaller number of energy states without releasing energy to their surroundings. Those energy states that carry energy away from the subsystem of molecules into the surrounding environment are the reason why total entropy increases despite the fact that the entropy of the subsystem decreases [8].

In other words, there will be a difference in entropy between the cell in a 20-year-old individual and the cell in the same individual at 80 years old. Consequently, in the following, we will use a hypothetical example that allows us to calculate the probability that a single one of this individual's cells, $\text{Cell}_{\text{old}}$, will undergo a spontaneous return to the state or order it had 60 years ago, when the individual was 20 years old and the cell, $\text{Cell}_{\text{young}}$, presumably was at its functional prime.

We will therefore begin with some estimates of the necessary multiplicity reduction. We will first study the reduction in multiplicity, that is, the number of accessible microstates, associated with constructing a single human cell from the beginning. We will then picture a series of more complex versions of this cell at 60-year intervals, going back over its 80-year history of development. Each newer version is somewhat more probable than its previous version, and the product of these multiplicity reductions should be sufficient to account for the necessary multiplicity reduction. The entropy reduction associated with the assembly of complex structures from simpler molecules comes in a variety of forms. To ease the calculations, we will evaluate just one part of this process, namely, the assembly of proteins from their building blocks, the amino acids.

[2] Assuming that a typical human epithelial cell is spherical, the volume will be estimated by $V = (4/3) \times \pi \times r^3 = (4/3) \times \pi \times (12/2)^3 = 904.8 \text{ mm}^3$ or $9.05 \times 10^{-9}$ cm³. The proteins are the main building blocks of the cell, taking up almost 20 % of a eukaryotic cell’s mass. Assuming that the density of the cell is that of water, (in fact, the density is closer to 1.3 times that of water, but that will be neglected here), the total mass of protein is $1.81 \times 10^{-9}$ g. The molecular weights of amino acids are roughly 150 g/mol, so the total number of amino acids in the proteins is approximately $7.24 \times 10^{12}$. An average protein is made of 300 amino acids, giving $2.4 \times 10^9$ proteins in the somatic cell. These are of course very rough estimates which would vary from cell to cell.

A single human cell contains approximately $2.4 \times 10^9$ proteins, a number that does not refer to the number of specific types of protein but rather to the total number of proteins in the somatic cell (see footnote 2). We will calculate the multiplicity cost of constructing all of these proteins by first estimating the multiplicity cost of constructing just one protein. Since we will ignore other processes, such as the synthesis of the amino acids in the first place, the construction of other polymers, etc., we will also underestimate the necessary multiplicity reduction.

Assume that the difference between $Cell_{old}$ today and $Cell_{young}$ 60 years ago is that one extra deformed protein has been created somewhere inside the cell (not one ***type*** of protein, but one solitary molecule). Assume that the protein contains 300 amino acids, which is roughly the average size of a protein molecule, and that those amino acids were already present in $Cell_{young}$, so that all that took place was that this protein was assembled one amino acid at a time. At each step, we must take an amino acid that was existing independently in the cell and put it in a specific position relative to the other amino acids that have already been constructed.

Before the protein was assembled, the individual amino acids could have been almost anywhere in the cell, but afterwards, they have to be in this specific order. That results in a huge reduction in multiplicity. A rough estimate of the multiplicity reduction is basically the degree to which the multiplicity of a solution of amino acids in fluid decreases when 299 amino acids are taken out of it (since those 299 have to be placed at specific locations relative to the 300th). This estimate can be calculated as follows

$$e^{-299\,\mu/kT}, \qquad (8)$$

where μ represents the chemical potential (that is, the Gibbs free energy per particle) of an amino acid [11]. Using the standard equation for an ideal gas

$$\mu = -T\{k[\ln(V\,(\frac{4\pi mU}{3h^2})^{3/2}) - \ln N^{5/2} + \frac{5}{2}] - Nk \cdot \frac{5}{2}\frac{1}{N}\} \qquad (9)$$

$$\mu = -kT\,\ln[\frac{V}{N}\,(\frac{2\pi mkT}{h^2})^{3/2}]$$

allows us to estimate the chemical potential [11]. If we make the most conservative assumptions possible, we find $\mu/kT \sim 10$, which means that assembling that one protein diminishes the multiplicity of the cell (that is, makes it more improbable) by a factor of about

$$e^{2990} \approx 10^{10^3}. \qquad (10)$$

That calculation is based on a number of simplistic assumptions, but a refinement of them will not change the fact that multiplicity changes are given by exponentially huge factors in systems like this. Generically, anything we do to a solitary molecule results in multiplicity changes given by $e^{-\mu/kT}$, where μ is always approximately of order eV, and kT is just approximately 0.025 eV [11].

If we wish to calculate the construction of all the proteins in the cell, then a slightly different approach than above can be adopted. Once again, we have that if the building blocks of the proteins were previously in a dilute solution in the cell, then the multiplicity reduction due to each of these steps is approximately $n_Q/n$, where n is the number density of amino acids and $n_Q$ is the density at which the amino acids would reach quantum degeneracy. To realize this, imagine that there are s amino acids in solution, with N available quantum states, where non-degeneracy means that N >> s. The multiplicity is $\Omega(s) = \binom{N}{s}$. Taking one amino acid out of solution causes the multiplicity to decrease by a factor Ω(s)/Ω(s - 1) = (N - s + 1)/s ≈ N/s = $n_Q$/n. This is definitely a large factor, demonstrating that amino acids in a cell are far from degenerate. To construct a protein with $N_a$ amino acids, it will be necessary to repeat this process $N_a$ -1 times, resulting in the exponentially huge number

$$\frac{\Omega_i}{\Omega_f} \sim (\frac{n_Q}{n})^{Na\,-\,1}. \tag{11}$$

We have again that if $n_Q$/n = 10 and $N_a$ = 300, the multiplicity ratio is then

$$\frac{\Omega_i}{\Omega_f} \sim 10^{299} \tag{12}$$

for the assembly of one specific protein molecule. Using this estimate for the multiplicity change associated with the assembly of one specific protein, we estimate the multiplicity reduction required to construct all of the proteins in the cell to be given as follows

$$\sim (10^{299})^{2.4 \text{ x } 10^9} \sim 10^{10^{12}}\,, \tag{13}$$

Assuming that 80 years (or 2.5 x $10^9$ seconds) of biological development were necessary to reach this, we would require a multiplicity reduction given as follows

$$(10^{10^{12}})^{1/2.5 \text{ x } 10^9} = 10^{400} \tag{14}$$

each second. This number represents the improbability, that is, that the desired young cell is $10^{400}$ times more improbable than the old cell. All of the previous estimates are of course very rough. For instance, they ignore the entropy changes due to the energy absorbed or emitted during the creation of chemical bonds. To include, we can pay attention to the fact that if, for instance, one chemical bond is broken, then an energy of roughly E = 1 eV is absorbed from the system, reducing the multiplicity with approximately $10^{17}$ at biological temperatures. The multiplicity change associated with a chemical reaction is $e^{\mu/kT}$. The chemical potential μ in a chemical reaction is usually of order 1 eV or more, indicating multiplicity changes of order $e^{40} \approx 10^{17}$ for each bond formed or broken at typical biological temperatures [5]. Changes always result in reductions in multiplicity. Since a great number of chemical bonds must be created in assembling each molecule, the resulting factor will once more be exponentially large, as above.

We can now enquire, what is the change in the entropy of a eukaryotic somatic cell due to changes over time? Assume that an individual cell, $Cell_{old}$, from a human being is $10^{400}$ times “more probable” than the same cell, $Cell_{young}$, was 60 years ago. That is, this individual cell is $10^{400}$ times more disordered than its previous self, which in turn was more precisely ordered. Therefore, if $\Omega_f$ is the number of microstates consistent with the specification at the present-time, less-improved, more-probable cell, and $\Omega_i$ is the number of microstates consistent with the specification of the cell 60 years ago, then this can be represented by

$$\Omega_f = 10^{-400}\,\Omega_i, \tag{15}$$

where $\Omega_i$ and $\Omega_f$ represent the initial and final cell. The statistical definition of entropy in the microcanonical ensemble is given by

$$S = k_B \ln \Omega, \tag{16}$$

where $k_B$ = 1.38 x $10^{-23}$ J/K is the Boltzmann constant [8]. The corresponding difference in entropy of $Cell_{old}$ of today versus $Cell_{young}$ 60 years ago is

$$S_f - S_i = k_B \ln \Omega_f - k_B \ln \Omega_i \quad (17)$$

$$= k_B \ln(\frac{\Omega_f}{\Omega_i}) = k_B \ln (10^{-400}) = - 127.1 \text{ x } 10^{-880} \text{ J/K}.$$

This is the entropy change over 60 years. Sixty years is 1.89 x $10^9$ seconds, and so the rate of entropy change per second getting from the past version of the cell to the present day cell is

$$\frac{S_f - S_i}{S_{years}} \quad (18)$$

$$= \frac{k_B ln(10^{-400})}{S_{years}} = -67.2 \text{ x } 10^{-889} \text{ J/K/sec.}$$

This is the entropy change for a single cell. However, perhaps there is a problem in identifying the reduction in the number of microstates from one generation (strictly speaking, it is not the same cell throughout the decades but rather its progeny) to the next because it is not exactly the same internal degrees of freedom of the proteins being rearranged, which are certainly fewer than those of the free amino acids. What we perhaps should be more aware of in this regard is the fact that eukaryotic cells are complex entities. It would therefore seem reasonable to speculate if whether it is possible to adjust that part of the estimates. However, as it stands, Equations (15) through (18) use the Boltzmann equation to find the change in entropy of a somatic cell, and this effectively takes the ratio of the number of energy states at the end to the number of energy states at the start of the interval of time. Therefore, it does not matter whether entropy refers to order or to the number of microstates. If order is proportional to the number of microstates, this proportionality cancels, meaning that despite the fact that the cell is a complex collection of molecules, in this instance, we do not need to know what the connection is. The above result is inserted as in Equation (7), and we now see that the probability of a cell being spontaneously rejuvenated is expressed by

$$\text{Cell}_{\text{Rejuvenated}} = (10^3)^{-67.2 \text{ x } 10^{-889}} \quad (19)$$

$$= 10^{-202 \text{ x } 10^{-889}}$$

Finally, we see that since an average human body contains approximately 10 trillion ($10^{13}$) cells [1], we can determine that the probability of all cells, i.e., a human being, undergoing spontaneously rejuvenation is expressed by

$$\text{Human}_{\text{Rejuvenated}} = 10^{-202 \text{ x } 10^{-889}} \text{ x } 10^{13} \quad (20)$$

$$= 10^{-2626 \text{ x } 10^{-889}}.$$

These numbers can without exaggeration be well described as overwhelming. The spontaneous rejuvenation of a cell turns out to involve movement from an almost certain situation to one that is unimaginably improbable. With numbers as large as the number of polymers in a cell, the probability that they would all move back together in a beneficial order and location as in previous times is incredibly small. The resulting probability is as close to zero as one can practically get. However, it is still a nonzero probability.

**DISCUSSION**

The second law of thermodynamics does not state that a cell must head toward a high entropy state and stay that way, it merely states that this is the most probable outcome in the future. Therefore, strictly speaking, it is entirely possible for entropy to spontaneously decrease, it is just that on a macroscopic scale, this is so improbable that the chance of this happening is usually unobservable in all practical situations. An important insight of statistical mechanics is that probability ratios for macroscopic systems are usually exponentially large quantities. Even small changes in a macroscopic system as large as a cell produce exponentially large changes in the multiplicity [10]. We have applied these insights to a cell and estimated the theoretical probability of its spontaneous return to a younger state. As we have observed, the probability of a cell returning to a previous, more ordered state is expressed by P = $10^{-202 \text{ x } 10^{-889}}$. The probability of all cells in the body undergoing this return is expressed by P = $10^{-2626 \text{ x } 10^{-889}}$. This is a large number, far exceeding the age of the universe. Unless you happen to be the luckiest individual in the known universe, counting on experiencing this event spontaneously, in yourself or another person, is not realistic, to say the least.

The previous section made use of a number of assumptions and simplifications. The calculations performed assume that all the different events in a cell are independent of each other, which might affect the probability being calculated for the part that is studied. Further, if we had considered more than just the formation of proteins, such as DNA, the result would in turn have been larger, resulting in a more accurate estimate. However, since water takes up almost 70% of the mass of a somatic cell, and proteins take up almost 20%, it seems fair to base the estimate solely on these (see footnote 2). Additionally, the calculations has included all the proteins in the cell, but a more accurate estimate would have been a lower number of proteins since it is unlikely that all the cell's proteins would be deformed and require return to a previous state. However, the goal was only to provide a rough estimate of what it would take for a cell to undergo a spontaneous return to a younger and fitter state of low entropy, which was enough to result in a large number of proteins alone.

However, a number of factors come into play that may suggest that such an event is not always given at such numbers. There are situations where order can arise relatively easily in spite of the large number of molecules involved. A good example is to take 500 of one type of atom and 500 of another atom and arrange them in a 10 x 10 x 10 cube. What are the chances of them arranging themselves in a sequence in which the two types of atoms neatly alternate? That chance would be $2^{500}$. However, if these atoms are sodium and chloride, this would be a piece of salt crystal. Even a cube with 100 atoms on a side (superficial change $2^{500,000}$) would not demand much. Of course, a cell is not as rigid and simple as a salt crystal, but it is considerably more structured than a mole of molecules in a gas container, and structure and the number of molecules do not always go together.

A classical definition of the second law states that during any reaction, the entropy of the universe will increase. However, the entropy of any local part of the universe can decrease with time, so long as that decrease is compensated by a bigger increase in a different part of the universe [10]. Does the concept of compensation make sense in the framework of thermal entropy? Yes, it certainly does! We can actually conduct an experiment illustrating that. Prepare a cup of hot water and put an ice cube in it. The water will cool down, and its entropy will fall. The former process could not have happened spontaneously: the decrease in entropy is much too large to permit this process to happen on its own. What permits this very improbable event to happen is a compensation. The ice cube is warming up. Despite the fact that the water, considered in isolation, is entering a much less likely state, the whole system (water and ice) is heading to a more likely state when the temperatures of water and ice move toward an equilibrium. As the water cools, it transfers an amount of heat Q to the ice cube, and the entropy of water decreases by $Q/T_w$. Releasing 1J of heat at room temperature $T_w$ = 300 K reduces the water entropy by $S_w$ = 1/300 J/K, which means that the number of microstates accessible to water decreases by a factor

$$\exp(\frac{S_W}{k_B}) = \exp(2.41 \times 10^{20}). \quad (20)$$

This is a large reduction of the phase space. There is almost no way this can occur on its own. Therefore, how is it possible? The answer to that question is that it comes from the compensation occurring in the ice. Having received Q = 1 J of heat at a lower temperature $T_i$ = 265 K, the ice cube increases its entropy by $S_i = Q/T_i$ = 1/265 J/K. The number of microstates accessible to it therefore goes up by

$$\exp(\frac{S_i}{k_B}) = \exp(2.74 \times 10^{20}). \quad (21)$$

The increase in the number of microstates of ice is also incredible large. The increase is much larger than the reduction factor for the number of microstates accessible to water. The combined system, water and ice, obtains access to more microstates than before. Their number is increased by

$$\exp(\frac{(S_i - S_w)}{k_B}) = \exp(0.31 \times 10^{20}). \quad (22)$$

Therefore, compensation obviously works regarding thermal entropy. It would be highly unlikely for a cup of water at room temperature to spontaneously cool down by even a fraction of a degree. The second law makes that highly improbable. However, if the corresponding entropy decrease is compensated by an equal or greater entropy increase in a cube of ice, then the process is permitted to occur, which demonstrates that a decrease in entropy is possible, even in spite of seemingly great odds.

The previous discussion also points to an important insight. The second law of thermodynamics states that entropy almost never decreases in closed systems. In open systems (systems with energy flows running through them), it clearly does; otherwise, there would be no inventions such as refrigerators, which makes an extremely interesting point: virtually nothing in biology happens under equilibrium conditions; equilibrium is basically death. To achieve anything (to make something change), we usually have to perform thermodynamic work, that is, we have to use free energy, which necessitates running other processes that release a greater amount of quantity of free energy, and couple these two processes together. Therefore, none of these are equilibrium processes [8].

Spontaneous order can emerge in open systems because of their ability to build their order by dissipating potentials in their environments. Schröedinger proposed the concept of living beings being streams of order that are allowed to exist away from equilibrium because they feed off "negentropy" (potentials) in their environments. As long as living systems produce entropy (or minimize potentials) at a sufficiently fast rate to compensate for their own internal ordering (their development and maintenance away from equilibrium, that is, their own internal entropy reduction), then the balance equation of the second law will not experience improbable events [12]. Living cells are open systems and counteract entropy for a limited time by feeding on free energy in food harvested from the surroundings [8]. A cell continuously transports specific chemicals in and out across its outer membrane to maintain homeostasis. These processes use energy that the cell abstracts from the environment. In cells, energy imported from outside powers a number of repair mechanisms that continually operate.

The overall effect is an increase in the entropy of the system, comprising the environment and the cell. Therefore, the cell counteracts and delays the effects of the second law for a long period in the organism [4]. Again, we see that the probability of a decrease in entropy can in fact be lower than anticipated. In all, it is possible that the probability of the molecules of an aging cell reverse into an earlier configuration, like the one when the cell was full of life would require a lower probability than one would otherwise expect.

Since the statistical interpretation of thermodynamic phenomena is a notion of probability, and it is well known that a strong energy input can force a system far away from equilibrium, then perhaps it could be reasonable to speculate, whether the said probability could be changed, i.e. could it theoretically be possible to manipulate probability and device a brilliant method to bring about the ability to increase these probabilities in our favor, and induce an otherwise incredibly unlikely event to happen as often as we please. It is however very hard to imagine ***what*** we could do to a system as large and complex as a somatic cell, which would change the number of available microstates, and bring this opportunity within our reach. But the fact remains that the possibility of an event occurring depends upon the event probability. And from this point of view, the aging process is not naturally inevitable and irreversible; the theoretical possibility of rejuvenation does exist.

It is important to note that even if the theory of the aging process of Toussaint et al. does turn out to be inadequate, the statistical reversibility of the aging process still applies. Deterioration of a cell's physiology over time is clearly a given, as evidenced by e.g. free radicals and a common characteristic of all aging process models, decay. The notion that the aging process can never really be fought is not correct, at least not from a theoretical point of view. Whether it will ever become a practical reality is another matter.